\documentclass[a4paper,12pt, epsfig]{article}
\pdfoutput=1
\usepackage{epsfig}
\usepackage{epstopdf}
\usepackage{graphicx}
\usepackage{ifthen}
\usepackage{feynmp}
\usepackage{amsmath}
\usepackage[psamsfonts]{amssymb}
\usepackage{euscript}

\usepackage{latexsym}
\usepackage[arrow,matrix,curve]{xy}

\newskip\humongous \humongous=0pt plus 1000pt minus 1000pt

\newif\ifdtup

\def\,{\hspace{-.1cm}}
\def\hsp{,\hspace{.7cm}}

\def\fc#1#2 {\frac{n}{q}#1\frac{n}{q}#2}

\def\vpp{V^{\prime\prime}}
\def\vppp{V^{\prime\prime\prime}}

\renewcommand{\cos}{\textrm{cos}}
\renewcommand{\sin}{\textrm{sin}}

\renewcommand{\sinh}{\textrm{sinh}}
\renewcommand{\cosh}{\textrm{cosh}}
\renewcommand{\tanh}{\textrm{tanh}}
\newcommand{\sech}{\textrm{sech}}
\newcommand{\csch}{\textrm{csch}}

\def\exp#1{\hbox{\rm exp}\left(#1\right)}

\renewcommand{\theequation}{\arabic{section}.\arabic{equation}}
\renewcommand{\(}{\begin{equation}}
\renewcommand{\)}{end{equation} \vspace{-.05in}\linebreak}

\newcounter{saveeqn}
\newcounter{savealpheqn}

\newcommand{\alpheqn}{\setcounter{saveeqn}{\value{equation}}%
  \stepcounter{saveeqn}\setcounter{equation}{0}%
  \renewcommand{\theequation}{\mbox{\arabic{section}.\arabic{saveeqn}
\alph{equation}}}
  \renewcommand{\)}{\end{equation}}}
\def\part#1{\frac{\partial}{\partial{#1}}}%
\def\group#1{\refstepcounter{equation}\setcounter{saveeqn}
 {\value{equation}}%
  \label{#1}\setcounter{equation}{0}%
\renewcommand{\theequation}{\mbox{\arabic{section}.\arabic{saveeqn}
\alph{equation}}}
  \renewcommand{\)}{\end{equation}}}
\newcommand{\reseteqn}{\setcounter{equation}{\value{saveeqn}}%
  \renewcommand{\theequation}{\arabic{section}.\arabic{equation}}%
  \renewcommand{\)}{\end{equation}}}

\newcommand{\aalpheqn}{\setcounter{saveeqn}{\value{equation}}%
  \stepcounter{saveeqn}\setcounter{equation}{0}%
  \renewcommand{\theequation}{\mbox{
        \Alph{subsection}.\arabic{saveeqn}\alph{equation}}}
   \renewcommand{\)}{\end{equation}}}
\newcommand{\areseteqn}{\setcounter{equation}{\value{saveeqn}}%
  \renewcommand{\theequation}{\Alph{subsection}.\arabic{equation}}%
  \renewcommand{\)}{\end{equation}}}

\renewcommand{\thefootnote}{\alph{footnote}}
\renewcommand{\(}{\begin{equation}}
\renewcommand{\)}{\end{equation}}
\newcommand{\ba}{\begin{eqnarray}}
\newcommand{\ea}{\end{eqnarray}}
\newcommand{\cbp}{\mathop{\vtop{\ialign{##\crcr
   $\hfil\displaystyle{}\hfil$\crcr\noalign{\kern-13pt\nointerlineskip}
   \BIG{)}\hskip0pt\crcr\noalign{\kern3pt}}}}}
\newcommand{\pa}{\mathop{\vtop{\ialign{##\crcr

$\hfil\displaystyle{\oplus}\hfil$\crcr\noalign{\kern+1pt\nointerlineskip
}
   \hspace{.08in}$^{\alpha=0}$\hskip6pt\crcr\noalign{\kern3pt}}}}}
\renewcommand{\hsp}{,\hspace{.3in}}
\newcommand{\p}{^\prime}

\newcommand{\Z}{\ensuremath{\mathbb Z}}

\catcode`\@=11
\def\vereq#1#2{\lower3pt\vbox{\baselineskip1.5pt \lineskip1.5pt
\ialign{$\m@th#1\hfill##\hfil$\crcr#2\crcr\sim\crcr}}}
\catcode`\@=12

\renewcommand{\(}{\begin{equation}}
\renewcommand{\)}{\end{equation}}

\def\pin#1{\int \frac{d#1}{2\pi}}
\def\pink#1{\int \frac{d^{#1}k}{(2\pi)^{#1}}}
\def\sq#1#2{\sqrt{\frac{\omega_{#1}}{\omega_{#2}}}}
\def\bd#1{b^\dag_{k_{#1}}}
\def\bm#1{b_{-k_{#1}}}

\def\df{\mathcal{D}_f}

\newcommand{\beas}{\begin{eqnarray*}}
\newcommand{\eeas}{\end{eqnarray*}}

\newcommand{\bquo}{\begin{quote}}
\newcommand{\enqu}{\end{quote}}

\renewcommand{\Z}{{\mathbb Z}}

\def\ch{{\mathcal{H}}}
\def\co{{\mathcal{O}}}

\newcommand{\beq}{\begin{equation}}
\newcommand{\eeq}{\end{equation}}
\newcommand{\bea}{\begin{eqnarray}}
\newcommand{\eea}{\end{eqnarray}}

\newskip\humongous \humongous=0pt plus 1000pt minus 1000pt

\newif\ifdtup

\catcode`\@=11

\@addtoreset{equation}{section}

\def\@normalsize{\@setsize\normalsize{15pt}\xiipt\@xiipt
\abovedisplayskip 14pt plus3pt minus3pt%
\belowdisplayskip \abovedisplayskip
\abovedisplayshortskip \z@ plus3pt%
\belowdisplayshortskip 7pt plus3.5pt minus0pt}

\def\small{\@setsize\small{13.6pt}\xipt\@xipt
\abovedisplayskip 13pt plus3pt minus3pt%
\belowdisplayskip \abovedisplayskip
\abovedisplayshortskip \z@ plus3pt%
\belowdisplayshortskip 7pt plus3.5pt minus0pt
\def\@listi{\parsep 4.5pt plus 2pt minus 1pt
      \itemsep \parsep
      \topsep 9pt plus 3pt minus 3pt}}

\relax

\catcode`@=12

\catcode`\@=11

\def\section{\@startsection{section}{1}{\z@}{3.5ex plus 1ex minus  .2ex}{2.3ex plus .2ex}{\large\bf}}

\def\thesection{\arabic{section}}
\def\thesubsection{\arabic{section}.\arabic{subsection}}

\def\appendix{\setcounter{section}{0}
 \def\thesection{Appendix \Alph{section}}
 \def\thesubsection{\Alph{section}.\arabic{subsection}}
 \def\theequation{\Alph{section}.\arabic{equation}}}
\renewcommand{\theequation}{\arabic{section}.\arabic{equation}}

\begin{document}
\def\thefootnote{\fnsymbol{footnote}}
\def\thetitle{Constructing Quantum Soliton States Despite Zero Modes}
\def\autone{Jarah Evslin}
\def\affa{Institute of Modern Physics, NanChangLu 509, Lanzhou 730000, China}
\def\affb{University of the Chinese Academy of Sciences, YuQuanLu 19A, Beijing 100049, China}

\begin{center}
{\large {\bf \thetitle}}

\bigskip

\bigskip

{\large \noindent  \autone{${}^{1,2}$} \footnote{jarah@impcas.ac.cn}}

\vskip.7cm

1) \affa\\
2) \affb\\

\end{center}

\begin{abstract}
\noindent
In classical Lorentz-invariant field theories, localized soliton solutions necessarily break translation symmetry.  In the corresponding quantum field theories, the position is quantized and, if the theory is not compactified, they have continuous spectra.  It has long been appreciated that ordinary perturbation theory is not applicable to continuum states.  Here we argue that, as the Hamiltonian and momentum operators commute, the soliton ground state can nonetheless be found in perturbation theory if one first imposes that the total momentum vanishes.  As an illustration, we find the subleading quantum correction to the ground state of the Sine-Gordon soliton.


\end{abstract}

%
\setcounter{footnote}{0}
\renewcommand{\thefootnote}{\arabic{footnote}}

\section{Introduction}
What is a quantum soliton?  In a classical theory, a soliton is a solution of the classical equations of motion with certain properties.  In a quantum theory, in the weak coupling limit, it is a coherent state defined entirely in terms of that classical solution \cite{hepp,sato}.  At small but finite coupling, solitons can be described by a semiclassical expansion about this coherent state \cite{taylor78}.  At strong coupling this expansion is generally meaningless\footnote{Even at weak coupling,  quantum corrections may lead to a violation of Derrick's theorem \cite{delfino,davies}.} and so the connection to the classical solution is elusive.  As a result, it is hard to see how a quantum soliton may be defined at strong coupling.  Yet there is plenty of evidence that quantum solitons at strong coupling are interesting and important, for example in the strongly coupled Sine-Gordon theory they become the fundamental fermions in the massive Thirring model \cite{colemansg,mandelstamsol}.  Also in $\mathcal{N}=2$ superQCD, softly broken to $\mathcal{N}=1$, a monopole condenses leading to confinement \cite{sw2}.  So what is a quantum soliton at strong coupling, where the semiclassical link to the classical solution is missing?  In the above two examples, a clear definition was provided respectively by integrability and by supersymmetry, but is there one in general?  It is our hope that an answer to these questions may shed led light on the ultimate questions: Just why is this superQCD monopole tachyonic?  And does the same mechanism \cite{thooftconf,mandelconf} work in real world QCD?

To answer these questions, our approach will be to follow the Sine-Gordon soliton, and eventually its supersymmetric avatar, as far into the quantum regime as we can.  Our approach is to use the Schrodinger picture of quantum field theory, where states exist on fixed time slices and operators are timeless.  This formalism has the advantage that the soliton and vacuum state are treated as two eigenstates of the same Hamiltonian, thus removing an old ambiguity\footnote{Another proposed solution, closer to the original approach, can be found in Ref.~\cite{rebsol}.} in the traditional approach \cite{dhn2,rajaraman,physrept04} which was first noted in Ref.~\cite{rebhan}.  Also, the traditional path-integral approach yields soliton energies but not the states themselves \cite{dhn1}, whereas we hope that finding the monopole state in superQCD will shed light on the physical mechanism that makes it tachyonic.  

We are interested in the soliton ground state, which corresponds to a time-independent state, and so time completely disappears from our formalism.  At one loop, the Sine-Gordon soliton is described by a free Poschl-Teller theory \cite{rajaraman}.  Recently \cite{mestato} we explicitly found the Schrodinger picture state corresponding to this ground state.  The solution was hardly surprising, as the theory is a free quantum field theory and so a sum of quantum harmonic oscillators, the one-loop state is a squeezed state.

In this paper we will find the first quantum correction to this state, which is relevant for two loop calculations.  It is tempting to use naive perturbation theory for this task.  However there is a complication.  The classical solution has a center of mass.  In the quantum theory, this corresponds to a collective coordinate.  In principle, the Hilbert space includes all wave functions of this collective coordinate, for example the soliton can have any momentum and so the spectrum is continuous.  It has long been appreciated \cite{friedrichscont} that usual perturbation theory does not apply in this setting.  We will see a direct manifestation of this below when we try to invert the free Hamiltonian and find that the inverse is not uniquely defined within our perturbative expansion.  

We propose a solution to this problem\footnote{There is an analogous problem in the path integral approach, and there the projection onto fixed momentum states is indeed known to solve the problem \cite{callangross}.}.  In 1+1 dimensions, continuous symmetries cannot be spontaneously broken \cite{coleman2d}.  The soliton is the ground state of a Sine-Gordon theory subjected to certain nontrivial boundary conditions, and so the corresponding state must be translation invariant\footnote{In Ref.~\cite{coleman2d} there was a heuristic derivation followed by a rigorous derivation.  The heuristic derivation applies as is despite nontrivial boundary conditions because these boundary conditions do not remove the divergence in the two-point function.}.  Therefore we first restrict the Hilbert space to the space of translation-invariant states.  These states still have a continuous spectrum, resulting from the existence of oscillators with arbitrarily low frequencies, however the continuity resulting from the soliton momentum is now gone.  We will see that as a result the free Hamiltonian is invertible and we are able to find the first quantum correction to the squeezed state.

 We begin in Sec.~\ref{revsez} with a review of the results at one loop, concentrating on our approach.  We review the basic setup for our problem, the one-loop energy of the soliton ground state and also our solution for the state itself.  Next in Sec.~\ref{wicksez} we find the Schrodinger equation which must be solved for the leading correction to this solution.  We attempt to solve it using ordinary perturbation theory, however we find that the inverse of the free Hamiltonian, needed to find a solution, is ambiguous.  In Sec.~\ref{psez} we describe our solution to this problem.  We find the relevant translation operator and use it to construct a general solution for a translation-invariant state.  Finally in Sec.~\ref{solsez} we repeat our perturbative analysis, now restricting attention to translation-invariant states.  This time we successfully find a unique leading correction to the one-loop state in a semiclassical expansion.  The most important elements of our notation are summarized in Table~\ref{notab}.

\section{A Review of the One Loop Solution} \label{revsez}

\begin{table}
\begin{tabular}{|l|l|}
\hline
Operator&Description\\
\hline
$\phi(x)$&The real scalar field\\
$\pi(x)$&Conjugate momentum to $\phi(x)$\\
$a^\dag_p,\ a_p$&Creation and annihilation operators in plane wave basis\\
$b^\dag_k,\ b_k$&Creation and annihilation operators in Poschl-Teller/soliton basis\\
$\phi_0,\ \pi_0$&Zero mode of $\phi(x)$ and $\pi(x)$ in Poschl-Teller/soliton basis\\
$::_a,\ ::_b$&Normal ordering with respect to $a$ or $b$ operators respectively\\
$P,\ P^\prime$&Momentum operator in Sine-Gordon theory and $\df$ shifted theory\\
\hline
Hamiltonian&Description\\
\hline
$H$&The Sine-Gordon Hamiltonian\\
$H^\prime$&$H$ with $\phi(x)$ shifted by soliton solution $f(x)$\\
$H_2$&The Poschl-Teller Hamiltonian\\
$H_3$&The leading interaction term in $H^\prime$\\
\hline
Symbol&Description\\
\hline
$f(x)$&The classical soliton solution\\
$\df$&Operator that translates $\phi(x)$ by the classical soliton solution\\
$g_B(x)$&The soliton linearized translation mode\\
$g_k(x)$&Continuum perturbation about the soliton solution\\
$p,\ q,\ r$&Momentum\\
$k_i$&The analog of momentum for soliton perturbations\\
$\omega_k,\ \omega_p$&The frequency corresponding to $k$ or $p$\\
$\tilde{g}$&Inverse Fourier transform of $g$\\
$\hat{g}$&Fourier transform of $\tilde{g}/\omega$ \\
$I(x)$&The loop factor which appears in tadpole diagrams\\
\hline
State&Description\\
\hline
$|K\rangle$&Soliton ground state\\
$|\Omega\rangle$&True ground state\\
$\co|\Omega\rangle$&Translation of $|K\rangle$ by $\df^{-1}$\\
$|0\rangle_n$&$n$th order of semiclassical expansion of $\co|\Omega\rangle$\\
$|0\rangle_n^{(k)}$&As above, with $k$ powers of $\phi_0$\\
\hline

\end{tabular}
\caption{Summary of Notation}\label{notab}
\end{table}

\subsection{Sine-Gordon to Poschl-Teller}

Consider a real scalar field $\phi(x)$ and its canonical momentum $\pi(x)$ in 1+1 dimensions, in a theory with Hamiltonian
\beq
H=\int dx \ch(x) \hsp
\ch(x)=\frac{1}{2}:\pi(x)\pi(x):_a+\frac{1}{2}:\partial_x\phi(x)\partial_x\phi(x):_a+V[\phi(x)].\label{hsq}
\eeq
For concreteness we will consider the case of the Sine-Gordon theory
\beq
V[\phi(x)]=\frac{m^2}{\lambda}\left(1-:\cos(\sqrt{\lambda}\phi(x)):_a\right)
\eeq
but the generalization to other potentials will be straightforward.   The normal-ordering $::_a$ will be defined below.

The classical equations of motion following from this Hamiltonian admit a time-independent soliton solution
\beq
\phi(x,t)=f(x)=\frac{4}{\sqrt{\lambda}}\arctan{e^{mx}}. \label{feq}
\eeq
The combination $\lambda\hbar$ is dimensionless and so the semiclassical expansion is an expansion in $\lambda$, where we set $\hbar=1$.  However the $\lambda^{-1/2}$ in the classical solution $f(x)$ prevents naive perturbation theory from capturing these solitons.   

A perturbative expansion about the soliton solution can nonetheless be defined.  We will use the strategy of \cite{mekink,memassa} in which one first defines a new Hamiltonian $H\p$ via the similarity transformation
\beq
H\p=\df^{-1} H\df \label{sim}
\eeq
where we have defined the translation operator
\beq
\df={\rm{exp}}\left(-i\int dx f(x)\pi(x)\right) \label{df}
\eeq
which satisfies the identity \cite{mekink}
\beq
:F\left[\pi(x),\phi(x)\right]:_a\df=\df:F\left[\pi(x),\phi(x)+f(x)\right]:_a \label{fident}
\eeq
for any functional $F$.

The soliton ground state is
\beq
|K\rangle=\df \co|\Omega\rangle
\eeq
where $\co$ is equal to the identity plus quantum corrections and $|\Omega\rangle$ is the ground state of a vacuum sector.  One can easily check that
\beq
H\p\co|\Omega\rangle=E\co|\Omega\rangle \label{quick}
\eeq
where $E$ is the soliton rest mass.  The problem of finding the ground state $|K\rangle$ (or any other energy eigenstate) of the soliton sector is thus equivalent to finding an eigenstate $\co|\Omega\rangle$ of $H\p$, and so one may forget the original Hamiltonian $H$ and study $H\p$.

Using (\ref{fident}), the new Hamiltonian $H\p$ may be expanded
\beq
H\p=Q_0+\sum_{n=2}^\infty H_n
\eeq 
where
\beq
Q_0=\frac{8m}{\lambda}
\eeq
is the classical soliton mass, $H_2$ is the Poschl-Teller Hamiltonian
\beq
H_2=\frac{1}{2}\int dx\left[:\pi^2(x):_a+:\left(\partial_x\phi(x)\right)^2:_a+V^{\prime\prime}[f(x)]:\phi^2(x):_a\right] \label{pt}
\eeq
and the interaction terms are
\beq
H_n=\frac{1}{n!}\int dx V^{(n)}[f(x)]:\phi^n(x):_a\hsp n>2 \label{hpexp}
\eeq
where $V^{(n)}$ is the $n$th derivative of the potential $V[\phi]$.  The term $H_n$ is proportional to $\lambda^{n/2-1}$ and so one may attempt to solve (\ref{quick}) perturbatively, keeping as many terms as are needed at each order.  The classical energy is of order $\lambda^{-1}$ and so the $m$-loop energy is of order $\lambda^{m-1}$, and therefore uses all terms in $H\p$ up to $H_{2m}$.

\subsection{Poschl-Teller at One Loop}

In Ref.~\cite{sg} we solved Eq.~(\ref{quick}) at one loop, providing an explicit expression for $\co|\Omega\rangle$ at one loop in Ref.~\cite{mestato}.   In the remainder of this section we will review that solution.

 At one loop one need only consider $H_2$.  Its classical equations of motion admit solutions
 \beq
 \phi(x,t)=e^{-i\omega t}g(x)\hsp
V^{\prime\prime}[f(x)]g(x)=\omega^2g(x)+g^{\prime\prime}(x). \label{cleq}
 \eeq
These are the equations of motion for a Poschl-Teller potential and the solutions are well known \cite{flugge}.  There is one bound state solution $g_B(x)$, corresponding to the translation mode. Translation is a symmetry and so the corresponding frequency is $\omega_B=0$.  There are also continuum modes $g_k(x)$ with frequency $\omega_k$ where we fix the index $k$ by demanding that $\omega_k^2=m^2+k^2$ and we fix the sign of $k$ by demanding that at large $\pm x$ the solution reduce to the corresponding plane wave, albeit with a phase shift.  Had we considered instead the $\phi^4$ theory, there would also have been another bound state corresponding to a breather mode of the kink.  More general theories might also correspond to potentials which are not reflectionless, in which case we would need to consider combinations of right and left moving modes.
 
We will impose the normalization conditions
\beq
\int dx g_{k_1} (x) g^*_{k_2}(x)=2\pi \delta(k_1-k_2)\hsp
\int dx |g_{B}(x)|^2=1 \label{comp}
\eeq
and note the orthogonality
\beq
\int dx g_{k_1}(x)g_B^*(x)=0.
\eeq
The solutions satisfy
\beq
g_k^*(x)=g_{-k}(x)\hsp g_B^*(x)=g_B(x).
\eeq 
Although we will not need them, for completeness we will write the explicit forms of these solutions
\beq
g_k(x)=\frac{e^{-ikx}}{\sqrt{1+m^2/k^2}}\left(1-i\frac{m}{k}\tanh(m x)\right)\hsp
g_{B}(x)=\sqrt{\frac{m}{2}}\sech\left(mx\right).  \label{geq}
\eeq
We will also define the inverse Fourier transforms of these functions as 
\bea
\tilde{g}_B(p)&=&\int dx g_B(x) e^{ipx}=\frac{\pi}{\sqrt{2m}}\sech\left(\frac{\pi p}{2m}\right)\\
\tilde{g}_k(p)&=&\int dx g_k(x) e^{ipx}=\frac{2\pi }{\sqrt{1+m^2/k^2}}\delta(p-k)+\frac{\pi}{k\sqrt{1+m^2/k^2}}\csch\left(\frac{\pi (p-k)}{2m}\right).\nonumber
\eea

The functions $g_k(x)$ and $g_B(x)$ are in fact a complete basis of the set of functions, being a complete set of eigenvectors of the operator $\partial_x^2-\vpp[f(x)]$ and also as evidenced by the completeness relation
\beq
g_B(x)g_B(y)+\pin{k}g_k(x)g_{-k}(y)=\delta(x-y)
\eeq
or equivalently
\beq
\tilde{g}_B(p)\tilde{g}_B(q)+\pin{k}\tilde{g}_k(p)\tilde{g}_{-k}(q)=2\pi\delta(p+q).
\eeq
 
 As the functions $g$ are a basis of the set of functions, they can be used to expand the field $\phi(x)$ and its canonical momentum $\pi(x)$.  More precisely, there are two expansions of interest.  The usual expansion in terms of plane waves is
\bea
\phi(x)&=&\pin{p}\phi_p e^{-ipx}\hsp
\phi_p=\frac{1}{\sqrt{2\omega_p}}\left(a^\dag_p+a_{-p}\right) \label{osc}\\
 \pi(x)&=&\pin{p}\pi_p e^{-ipx}\hsp
\pi_p=i\sqrt{\frac{\omega_p}{2}}\left(a^\dag_p-a_{-p}\right)\hsp
\omega_p=\sqrt{m^2+p^2}
\nonumber
\eea
while the expansion in terms of Poschl-Teller eigenfunctions is 
\bea
\phi(x)&=&\phi_0 g_B(x)+\pin{k}\phi_k g_k(x)\hsp
\pi(x)=\pi_0 g_B(x)+\pin{k}\pi_k g_k(x)
\nonumber\\
\phi_k&=&\frac{1}{\sqrt{2\omega_k}}\left(b_k^\dag+b_{-k}\right)\hsp
\pi_k=i\sqrt{\frac{\omega_k}{2}}\left(b_k^\dag - b_{-k}\right). \label{pib}
\eea
We define two normal ordering prescriptions.  The operator $:O:_a$ will be ordered so that when decomposed in terms of $a^\dag$ and $a$, all $a^\dag$ are on the left.  The operator $:O:_b$ will be ordered so that when decomposed in terms of $\phi_0,\ \pi_0,\ b^\dag$\ and $b$, all $b^\dag$ and $\phi_0$ are on the left.  The Hamiltonian (\ref{hsq}) was defined in terms of $a$ normal ordering, and the mismatch between the two normal-ordering schemes is responsible for the one-loop correction to the mass \cite{mekink,sg}.  We will refer to $::_b$ as soliton normal ordering.

We will consistently use the index $k$ for the Poschl-Teller momentum, while $p$, $q$ and $r$ will be used for the true momentum.   This means, for example, that $\phi_p$ and $\phi_k$ are distinct operators, indeed they are coefficients of $\phi$ as expanded in distinct bases.  Sometimes it will be convenient to separate the bound and continuum parts of the fields
\bea
\phi_B(x)&=&\phi_0 g_B(x)\hsp
\phi_C(x)=\pin{k}\phi_k g_k(x)\\
\pi_B(x)&=&\pi_0 g_B(x)\hsp
\pi_C(x)=\pin{k}\pi_k g_k(x).\nonumber
\eea

As the plane waves and Poschl-Teller eigenfunctions are both complete bases of the space of functions, the above decompositions are easily inverted, one simply integrates $\phi(x)$ and $\pi(x)$ weighted by the complex conjugate of a basis function to arrive at the corresponding mode.  Therefore the canonical commutation relation $[\phi(x),\pi(y)]=i\delta(x-y)$ determines the algebra of the components
\beq
[a_p,a^\dag_q]=2\pi\delta(p-q)\hsp
[\phi_0,\pi_0]=i\hsp
[b_{k_1},b^\dag_{k_2}]=2\pi\delta(k_1-k_2) \label{alg}
\eeq
with other commutators within each decomposition vanishing as usual.

Composing the inverse of the $a$ decomposition with the $b$ decomposition, one obtains the Bogoliubov transform which relates them
\bea
a^\dag_p&=&a^\dag_{B,p}+a^\dag_{C,p}\hsp
a_p=a_{B,p}+a_{C,p}\label{bog}\\
a^\dag_{B,p}&=&\tilde{g}_{B}(p)\left[ \sqrt{\frac{\omega_p}{2}}\phi_0-\frac{i}{\sqrt{2\omega_p}}\pi_0\right]\hsp
a_{B,-p}=\tilde{g}_{B}(p)\left[ \sqrt{\frac{\omega_p}{2}}\phi_0+\frac{i}{\sqrt{2\omega_p}}\pi_0\right].\nonumber\\
a^\dag_{C,p}&=&\pin{k}\frac{\tilde{g}_k(p)}{2}\left(\frac{\omega_p+\omega_k}{\sqrt{\omega_p\omega_k}}b_k^\dag+\frac{\omega_p-\omega_k}{\sqrt{\omega_p\omega_k}}b_{-k}\right) \nonumber\\
a_{C,-p}&=&\pin{k}\frac{\tilde{g}_k(p)}{2}\left(\frac{\omega_p-\omega_k}{\sqrt{\omega_p\omega_k}}b_k^\dag+\frac{\omega_p+\omega_k}{\sqrt{\omega_p\omega_k}}b_{-k}\right)\nonumber.
\eea
Inserting (\ref{bog}) into the Poschl-Teller Hamiltonian (\ref{pt}) one obtains the one-loop Hamiltonian in terms of the $b$ oscillators
\beq
H_{2}=Q_1+ \pin{k}\omega_k b^\dag_k b_k+
\frac{\pi_0^2}{2} \label{hfin}
\eeq
where $Q_1$ is the one-loop correction to the soliton mass
\bea
Q_1&=&-\frac{1}{4}\pin{k}\pin{p}\frac{(\omega_p-\omega_k)^2}{\omega_p}\tilde{g}^2_{k}(p)
-\frac{1}{4}\pin{p}\omega_p\tilde{g}_{B}(p)\tilde{g}_{B}(p).\nonumber
\eea

The one-loop Hamiltonian (\ref{hfin}) is the sum of a free quantum-mechanical particle described by $\phi_0$ and $\pi_0$ and describing the center of mass motion of the soliton, with an infinite number of quantum harmonic oscillators labeled by the index $k$.  The one-loop ground state is thus the tensor product of the vacua of these various quantum mechanical sectors.  More precisely, if we decompose $\co|0\rangle$ using a semiclassical expansion
\beq
\co|\Omega\rangle=\sum_{n=0}^\infty |0\rangle_n \label{semi}
\eeq
where $|0\rangle_n$ is the contribution arising at $O(\lambda^{n/2})$ then at one-loop the ground state satisfies
\beq
b_k|0\rangle_0=\pi_0|0\rangle_0=0. \label{coeq}
\eeq
These conditions were solved in Ref.~\cite{mestato} to obtain the one-loop ground state $\df|0\rangle_0$, which we now recall.  A basis of states is given by the eigenvectors $|\Phi\rangle$ of the field $\phi(x)$
\beq
\phi(x)|\Phi\rangle=\Phi(x)|\Phi\rangle
\eeq  
where the eigenvalues are functions\footnote{Recall that a quantum field $\phi$ corresponds to one operator at each point $x$, and each of these operators has eigenvectors with eigenvalues.  Therefore an eigenvalue of $\phi$ is actually a choice of eigenvalue at every point $x$, or in other words a function $\Phi:x\mapsto\Phi(x)$.} $\Phi(x)$.  In terms of this basis, the state $|0\rangle_0$ is given by coefficients which are functionals $\Psi_0$ of the functions $\Phi(x)$
\bea
|0\rangle_0&=&\int D\Phi \Psi_0[\Phi] |\Phi\rangle\hsp
\Phi_k=\int dx \Phi(x) g^*_k(x)\nonumber
\\
\Psi_0[\Phi]&=&\exp{-\frac{1}{2}\pin{k}\Phi_k\omega_k\Phi_{-k}}\label{s00}
\eea
while the one-loop ground state $\df|0\rangle_0$ is given by
\bea
\df|0\rangle_0&=&\int D\Phi \Psi_K[\Phi] |\Phi\rangle\hsp
f_k=\int dx f(x) g^*_k(x)\nonumber
\\
\Psi_K[\Phi]&=&\exp{-\frac{1}{2}\pin{k}\left(\Phi_k-f_{k}\right)\omega_k\left(\Phi_{-k}-f_{-k}\right)}.
\eea
One sees that the one-loop ground state is a squeezed state.  Thus concludes our review.  The goal of this paper will be to find the correction $|0\rangle_1$.

\section{Soliton Normal Ordering the Interaction Terms} \label{wicksez}

\subsection{Setup}

At subleading order
\beq
\co|\Omega\rangle=|0\rangle_0+|0\rangle_1
\eeq
and so the Schrodinger equation (\ref{quick}) reduces to
\beq
H_2|0\rangle_0=Q_1|0\rangle_0\hsp
H_3|0\rangle_0=-(H_2-Q_1)|0\rangle_1. \label{seqs}
\eeq
In the previous section we reviewed the solution (\ref{s00}) of the first of these equations.  The goal of the rest of this note will be to solve the second.

In light of Eq.~(\ref{coeq}), it will be convenient to reexpress
\beq
H_3=\frac{1}{6}\int dx \vppp(x) :\phi^3(x):_a
\eeq
in terms of soliton normal ordered products $:O:_b$.   As $\phi_0$ and $b_k$ commute, we may decompose $:\phi^3(x):_a$ as
\beq
:\phi^3(x):_a=:\phi_B^3(x):_a+3:\phi_B^2(x):_a\phi_C(x)+3\phi_B(x):\phi^2_C(x):_a+:\phi_C^3(x):_a.
\eeq
We will calculate each of these terms in turn.  

\subsection{$n$-Point Functions}

The $a$ normal ordering is defined in terms of oscillators $a^\dag$ and $a$, therefore to evaluate these terms we first expand in terms of plane waves using (\ref{osc}), then the expressions are converted into $b^\dag$ and $b$ using (\ref{bog}) and using the commutators in (\ref{alg}) these are soliton normal ordered.  Terms with just one field are already normal ordered, so we only need to consider terms with two or three fields.  As bound and continuum fields commute with each other, we need only consider terms with two or three $\phi_B$ or two or three $\phi_C$.  

The simplest product is the square of the bound component of the field
\bea
:\phi^2_B(x):_a&=&\pin{p}\pin{q}\frac{e^{-ix(p+q)}}{\sqrt{4\omega_p\omega_q}}:\left(a^\dag_{B,p}+a_{B,-p}\right)\left(a^\dag_{B,q}+a_{B,-q}\right):_a\\
&=&\pin{p}\pin{q}\frac{e^{-ix(p+q)}}{\sqrt{4\omega_p\omega_q}}\left[a^\dag_{B,p}\left(a^\dag_{B,q}+a_{B,-q}\right)+\left(a^\dag_{B,q}+a_{B,-q}\right)a_{B,-p}\right]\nonumber\\
&=&\pin{p}\pin{q}\frac{e^{-ix(p+q)}}{\sqrt{4\omega_p\omega_q}}\tilde{g}_B(p)\tilde{g}_B(q)\nonumber\\
&&\times\left[\left(\sqrt{\omega_p}\phi_0-\frac{i}{\sqrt{\omega_p}}\pi_0\right)\sqrt{\omega_q}\phi_0+\sqrt{\omega_q}\phi_0\left(\sqrt{\omega_p}\phi_0+\frac{i}{\sqrt{\omega_p}}\pi_0\right)\right]\nonumber\\
&=&g^2_B(x) \phi_0^2-i\pin{p}\pin{q}\frac{e^{-ix(p+q)}}{\sqrt{4\omega_p\omega_q}}\tilde{g}_B(p)\tilde{g}_B(q)\sqrt{\frac{\omega_q}{\omega_p}}[\pi_0,\phi_0]\nonumber\\
&=&:\phi_B^2(x):_b-\pin{p}\pin{q}\frac{e^{-ix(p+q)}}{2\omega_p}\tilde{g}_B(p)\tilde{g}_B(q)\nonumber\\
&=&:\phi_B^2(x):_b-g_B(x)\hat{g}_B(x)\nonumber
\eea
where in the last line we introduced the shorthand notation
\beq
\hat{g}(x)=\pin{p}\frac{e^{-ipx}}{2\omega_p}\tilde{g}(p)
\eeq
which we will define both for the bound state function $g_B$ and also the continuum $g_k$.  Note that our answer resembles the usual Wick's theorem, with $1/(2\omega_p)$ the propagator arising from the contraction of two fields.

The square of the projection of the field $\phi$ onto the continuum is quite similar
\bea
:\phi^2_C(x):_a&=&\pin{p}\pin{q}\frac{e^{-ix(p+q)}}{\sqrt{4\omega_p\omega_q}}\left[a^\dag_{C,p}\left(a^\dag_{C,q}+a_{C,-q}\right)+\left(a^\dag_{C,q}+a_{C,-q}\right)a_{C,-p}\right]\nonumber\\
&=&\frac{1}{2}\pin{p}\pin{q}\frac{e^{-ix(p+q)}}{\sqrt{4\omega_p\omega_q}}\pink{2}\tilde{g}_{k_1}(p)\tilde{g}_{k_2}(q)\nonumber\\
&&\times\left[\left(\left[\sq{p}{k_1}+\sq{k_1}{p}\right]\bd{1}+\left[\sq{p}{k_1}-\sq{k_1}{p}\right]\bm{1}\right)\sq{q}{k_2}\left(\bd{2}+\bm{2}  \right)\right.\nonumber\\
&&+\left.\sq{q}{k_2}\left(\bd{2}+\bm{2}  \right) \left(\left[\sq{p}{k_1}-\sq{k_1}{p}\right]\bd{1}+\left[\sq{p}{k_1}+\sq{k_1}{p}\right]\bm{1}\right)    \right]\nonumber\\
&=&\frac{1}{2}\pin{p}\pin{q}\pink{2}\frac{e^{-ix(p+q)}}{\sqrt{4\omega_{k_1}\omega_{k_2}}}\tilde{g}_{k_1}(p)\tilde{g}_{k_2}(q)\nonumber\\
&&\times\left[2\left(\bd{1}\bd{2}+\bd{1}\bm{2}+\bd{2}\bm{1}+\bm{1}\bm{2}\right)\right.\nonumber\\
&&\left.+\left(1-\frac{\omega_{k_1}}{\omega_p}\right)\left([\bm{1},\bd{2}]+[\bm{2},\bd{1}]\right)\right]\nonumber\\
&=&\pink{2}g_{k_1}(x)g_{k_2}(x):\phi_{k_1}(x)\phi_{k_2}(x):_b    \nonumber\\
&&+\pin{p}\pin{q}\pin{k}e^{-ix(p+q)}\tilde{g}_{k}(p)\tilde{g}_{-k}(q)\left(\frac{1}{2\omega_k}-\frac{1}{2\omega_p}\right)
\nonumber\\
&=&:\phi_C^2(x):_b+\pin{k}\left(\frac{g_k(x)}{2\omega_k}-\hat{g}_k(x)\right)g_{-k}(x).\nonumber
\eea
We see that the Wick's theorem relating vacuum and soliton normal ordering, in the case of the continuum parts of the fields, replaces each contraction with $1/\omega_k-1/\omega_p$.  Intuitively the first term arises from the soliton normal ordering and the second from the vacuum normal ordering.  In the case of the bound parts of fields, which do not contain $b$ operators, only the second term appeared.  As these contraction terms will appear again in the three point functions, we will name them
\beq
I(x)=I_B(x)+I_C(x)\hsp
I_B(x)=-g_B(x)\hat{g}_B(x)\hsp
I_C(x)=\pin{k}\left(\frac{g_k(x)}{2\omega_k}-\hat{g}_k(x)\right)g_{-k}(x).
\eeq

Let us now calculate $I(x)$. 
\bea
I(x)&=&-g_B(x)\hat{g}_B(x)-\pin{k}g_{k}(x)\hat{g}_{-k}(x)+\pin{k}g_{k}(x)\frac{{g}_{-k}(x)}{2\omega_k}\\
&=&-\pin{q}e^{-iqx} \pin{p}\frac{e^{-ipx}}{2\omega_p} \left[\tilde{g}_B(q)\tilde{g}_B(p)+\pin{k}\tilde{g}_k(q)\tilde{g}_{-k}(p)\right]\nonumber\\
&&+\pin{k}\frac{1}{2\omega_k}g_{k}(x)\pin{p}e^{-ipx}\tilde{g}_{-k}(p)\nonumber\\
&=&-\pin{q}e^{-iqx} \pin{p}\frac{e^{-ipx}}{2\omega_p} 2\pi\delta(p+q)+\pin{k}\frac{1}{2\omega_k}g_{k}(x)\pin{p}e^{-ipx}\tilde{g}_{-k}(p)\nonumber\\
&=&\pin{p}\left[-\frac{1}{2\omega_p}+e^{-ipx}\pin{k}\frac{1}{2\omega_k}g_{k}(x)\tilde{g}_{-k}(p)\right].
\eea
It is finite.  Notice that the first two terms in the first line have condensed into a single term which is $x$-independent.  This term, on its own, would have been infinite after the $p$ integration.  However the third term ensures that the $p$ integral is finite.  Now if we take the derivative with respect to $x$, only the last term contributes
\beq
\partial_x I(x)=\pin{k}\frac{1}{2\omega_k}\partial_x\left|g_{k}(x)\right|^2. \label{di}
\eeq
This, together with the boundary condition that $I(x)$ vanishes as $x$ goes to infinity, completely determines $I(x)$.  Note that we have nowhere used the explicit form of the ground state, so we expect (\ref{di}) to hold to for any kink in a scalar theory with any potential, so long as the $k$ integral also includes a sum over breather modes.

Using
\beq
\left|g_{k}(x)\right|^2=1-\frac{m^2{\rm{sech}}^2(mx)}{\omega_k^2}
\eeq
one finds
\beq
I(x)=-\frac{\rm{sech}^2(mx)}{2\pi}.
\eeq




The calculations of the three point functions are quite similar to those of the two point functions.  First, for the bound part of the field
\bea
:\phi^3_B(x):_a&=&\pin{p}\pin{q}\pin{r}\frac{e^{-ix(p+q+r)}}{\sqrt{8\omega_p\omega_q\omega_r}}\\
&&\times\left[a^\dag_{B,p}\left(a^\dag_{B,q}\left(a^\dag_{B,r}+a_{B,-r}\right)+\left(a^\dag_{B,r}+a_{B,-r}\right)a_{B,-q}\right)\right.\nonumber\\
&&\left.+\left(a^\dag_{B,q}\left(a^\dag_{B,r}+a_{B,-r}\right)+\left(a^\dag_{B,r}+a_{B,-r}\right)a_{B,-q}\right)a_{B,-p}\right]\nonumber\\
&=&\pin{p}\pin{q}\pin{r}e^{-ix(p+q+r)}\tilde{g}_B(p)\tilde{g}_B(q)\tilde{g}_B(r)\left(\phi_0^3-3\frac{\phi_0}{2\omega_p}\right)\nonumber\\
&=&:\phi^3_B(x):_b+3I_B(x)\phi_B(x)\nonumber.
\eea
The interpretation in terms of Wick's theorem is clear, there are three contractions possible among the three factors of $\phi_B$, each yielding a factor of $I_B(x)$.  Finally we can compute
\bea
:\phi^3_C(x):_a&=&:\phi^3_C(x):_b+3\pin{p}\pin{q}\pin{r}\pink{2}e^{-ix(p+q+r)}\\
&&\times \tilde{g}_{k_1}(p)\tilde{g}_{-k_1}(q)\tilde{g}_{k_2}(r)\left(\frac{1}{2\omega_{k_1}}-\frac{1}{2\omega_p}\right)\frac{\bd{2}+\bm{2}}{\sqrt{2\omega_{k_2}}}\nonumber\\
&=&:\phi^3_C(x):_b+3I_C(x)\phi_C(x).\nonumber
\eea

Assembling our results, we can evaluate $H_3$ on the one-loop state $|0\rangle_0$
\bea
H_3|0\rangle_0&=&\left(A\phi_0^3+\pink{1}B_{k_1}\phi_0^2\frac{\bd{1}}{\sqrt{2\omega_{k_1}}}+\pink{2}C_{k_1k_2}\phi_0\frac{\bd{1}\bd{2}}{\sqrt{4\omega_{k_1}\omega_{k_2}}}+D\phi_0\right.\label{h3form}\\
&&\left.+\pink{3}E_{k_1k_2k_3}\frac{\bd{1}\bd{2}\bd{3}}{\sqrt{8\omega_{k_1}\omega_{k_2}\omega_{k_3}}}+\pink{1}F_{k_1}\frac{\bd{1}}{\sqrt{2\omega_{k_1}}}\right)|0\rangle_0\nonumber.
\eea
Adopting the shorthand
\beq
\vppp_{IJK}=\int dx \vppp[f(x)]g_I(x)g_J(x)g_K(x) \label{short}
\eeq
where the indices can be $B$ or $k_i$, we find
\bea
A&=&\frac{1}{6}\vppp_{BBB}=0\hsp
B_k=\frac{1}{2}\vppp_{BBk}\hsp
C_{k_1k_2}=\frac{1}{2}\vppp_{Bk_1k_2}\\
D&=&\frac{1}{2}\int dx \vppp[f(x)]g_B(x)I(x)=0\hsp
E_{k_1k_2k_3}=\frac{1}{6}\vppp_{k_1k_2k_3}\nonumber\\
F_k&=&\frac{1}{2}\int dx \vppp[f(x)]g_k(x)I(x).\nonumber
\eea
The constants $A$ and $D$ vanish because they are the integrals of products of even functions times $\vppp$, which is odd.

\subsection{The Problem}

Now we have the left hand side of the second Schrodinger equation in Eq.~(\ref{seqs}).  So can we solve it for the leading correction $|0\rangle_1$ to the soliton state?  To do this, we must invert $H_2$.  Intuitively this must be possible as $H_2$ is the sum of a square, which must be positive definite, and a series of harmonic oscillators, which are also positive definite.  As the soliton basis of operators consists of a canonical algebra $\phi_0$ and $\pi_0$ and also harmonic oscillators $b^\dag$ and $b$, the Hilbert space itself can be represented as a tensor product of a quantum mechanical wave function in $\phi_0$ and oscillator states.  Then the $\pi_0^2$ term in $H_2$ is $-\partial_{\phi_0}^2$, acting on these wave functions.  

Let us try a simple example.  Find the state $|\psi\rangle$ that satisfies
\beq
H_2|\psi\rangle=\bd{1}|0\rangle_0.
\eeq
Unfortunately there is more than one answer:
\beq
|\psi\rangle=\left(\frac{1}{\omega_{k_1}}+\beta \cos\left(\sqrt{2\omega_{k_1}}\phi_0\right)+\gamma\sin\left(\sqrt{2\omega_{k_1}}\phi_0\right) \right)\bd{1}|0\rangle_0 \label{due}
\eeq
for any numbers $\beta$ and $\gamma$.

What went wrong?  If we naively apply perturbation theory, we solve for $|0\rangle_1$ order by order in $\phi_0$.  But at any finite order, in fact any order greater than two,  this leads to a polynomial in $\phi_0$ and thus $\pi_0^2$ on the wave function is unbounded.  Indeed, the fact that $\pi_0^2$ is positive definite comes from the fact that it arose from a Hamiltonian consisting of squares, but this structure has been hidden by an integration by parts.  Thus the zero eigenvalues of $H_2$ acting on the $\beta$ and $\gamma$ terms in Eq.~(\ref{due}) are not obviously forbidden in perturbation theory.  The integration by parts cannot be undone when the wave function is a polynomial in $\phi_0$ because it diverges and so the boundary terms diverge.  Of course this divergence is fictitious, because the wave function is not really polynomial in $\phi_0$, that is simply the organization of the perturbation theory.  However this leaves us with the problem that in perturbation theory, $H_2$ does not seem to have a unique inverse and so one cannot solve for $|0\rangle_1$ without further inputs.

\noindent
{\bf Summary}: We found $H_3|0\rangle_0$ but we cannot uniquely invert $H_2$ to obtain $|0\rangle_1$ using Eq.~(\ref{seqs}).

\section{The Zero Momentum Sector} \label{psez}

\subsection{The Solution}

The problem with the invertibility of $H_2$ comes from the existence of the flat direction corresponding to translations of the soliton.  As the original Hamiltonian had a translation symmetry, this is an exact symmetry of the system and so of the ground state wave function.  There is also a continuous spectrum of states above it corresponding to small momenta for the soliton.  In general it is known \cite{vanhove1,vanhove2} that perturbation theory fails for continuous spectra because they lead to interesting physical effects, such as clouds, that are not captured by perturbation theory.

However in this case the flat direction corresponds to a symmetry which commutes with the Hamiltonian and, in particular, it is an exact symmetry of the ground state.   Thus the Hamiltonian does not mix states with different momenta.  The zero momentum states are a series of harmonic oscillators, each of which is gapped (although there is a limit as $k\rightarrow 0$ in which the gap becomes small).  As a result we do not expect the continuum to lead to any exotic physics.  On the contrary, if we first restrict to zero momentum states then we expect ordinary perturbation theory to be reliable.  We will see that the zero momentum condition itself is rather complicated and can only be solved in perturbation theory.  However it will be sufficient to first solve it at the desired order, and then perform perturbation theory on the restricted states at that order.  This will be our strategy\footnote{Another strategy has been employed at one loop in Ref.~\cite{sakitacc}.  We believe that our approach is more direct.}.

The momentum operator is
\beq
P=-\int dx :\pi(x) \partial_x \phi(x):_a=\pin{p} p a^\dag_p a_p.
\eeq
This commutes with the Sine-Gordon Hamiltonian $H$ in (\ref{hsq}).  However our perturbation theory is a decomposition of $H\p$, which was defined by the similarity transform (\ref{sim}).  Therefore $H\p$ does not commute with $P$, it is not translation invariant, instead it commutes with the similarity transform
\beq
[H\p,P\p]=0\hsp
P\p=\df^{-1}P\df=-\int dx :\pi(x) \partial_x (\phi(x)+f(x)):_a=-\pi_0/\alpha+P \label{pp}
\eeq
where we have defined the constant of proportionality $\alpha$ by
\beq
g_B(x)=\alpha f^\prime(x). \label{prop1}
\eeq
We note that
\beq
\frac{1}{\alpha^2}=\int dx f^{\prime 2}(x)
\eeq
is twice the kinetic energy term in Eq.~(\ref{hsq}) corresponding to the soliton solution and in fact is equal to the classical energy $Q_0$.  It can be directly calculated from  Eqs.~(\ref{feq}) and (\ref{geq})
\beq
\alpha=\sqrt\frac{\lambda}{8m}=\frac{1}{\sqrt{Q_0}}. \label{prop2}
\eeq

Now we are ready for the key step in our analysis.  The central observation is that, as the theory is translation invariant and translation symmetry cannot be spontaneously broken in 1+1 dimensions, the ground state of the soliton sector must also be translation invariant
\beq
0=P|K\rangle=P\df \co|\Omega\rangle=\df P\p\co|\Omega\rangle.
\eeq
Left multiplying by $\df^{-1}$ we find
\beq
P\p\co|\Omega\rangle=0. \label{ppinv}
\eeq
This condition can be expanded order by order using (\ref{semi}) and (\ref{pp}).  The leading term is
\beq
-\sqrt\frac{8m}{\lambda}\pi_0|0\rangle_0=0.
\eeq
This is satisfied already due to the definition of $|0\rangle_0$ in Eq.~(\ref{coeq}).  In this paper we are interested in the subleading contribution to the state.  It arises from the subleading term in~(\ref{ppinv})
\beq
P|0\rangle_0=\sqrt\frac{8m}{\lambda}\pi_0|0\rangle_1. \label{princ}
\eeq
Our strategy in this paper will be to first impose (\ref{princ}).  This will costrain $|0\rangle_1$ but not fix it entirely.  However we will see that it fixes it sufficiently so that $H_2$ can be inverted and so the Schrodinger equation (\ref{seqs}) can be solved. More generally, we claim the following.

\noindent
{\bf Claim:}{\it{ First impose momentum invariance on the ground state at a given order in $\lambda$ by solving Eq.~(\ref{ppinv}), expanded as described in Eqs.~(\ref{semi}) and (\ref{pp}).  Then the Schrodinger equation (\ref{quick}), expanded using (\ref{hpexp}), can be uniquely solved at the same order.}}

\subsection{The Momentum Operator}

To solve (\ref{princ}) we need to calculate the action of $P$ on $|0\rangle_0$.  It will be convenient to calculate $P$ in the soliton basis of operators $\phi_0,\ \pi_0,\ b^\dag$\ and $b$.  First note that
\bea
a^\dag_p a_p&=&\frac{1}{2}\tilde{g}_B(p)\tilde{g}_B(-p)\left(\omega_p\phi_0^2+\frac{1}{\omega_p}\pi_0^2+[\phi_0,\pi_0]\right)\\
&&+\frac{1}{2}\pink{1}\left[\left(\tilde{g}_B(-p)\tilde{g}_{k_1}(p)+\tilde{g}_B(p)\tilde{g}_{k_1}(-p)\right)\left(\omega_p\phi_0\phi_{k_1}+\frac{1}{\omega_p}\pi_0\pi_{k_1}\right)\right.\nonumber\\
&&\left. +\left(\tilde{g}_B(-p)\tilde{g}_{k_1}(p)-\tilde{g}_B(p)\tilde{g}_{k_1}(-p)\right)\left(i\pi_0\phi_k-i\phi_0\pi_k\right)\right]\nonumber\\
&&+\frac{1}{2}\pink{2}\tilde{g}_{k_1}(p)\tilde{g}_{k_2}(-p)\left[\omega_p\phi_{k_1}\phi_{k_2}+\frac{1}{\omega_p}\pi_{k_1}\pi_{k_2}+i\left(\phi_{k_1}\pi_{k_2}-\pi_{k_1}\phi_{k_2}\right)\right]\nonumber.
\eea
To obtain $P$, we need to integrate over $p$, weighted by $p$.  This eliminates all terms in $a^\dag_p a_p$ which are even in $p$, including all terms which include only bound state fields or scalars, leaving only terms which are products of a $\phi$ with a $\pi$
\bea
P&=&\pin{p}p a^\dag_p a_p\\
&=&\frac{i}{2}\pin{p}p\left[\pink{1}\left(\tilde{g}_B(-p)\tilde{g}_{k_1}(p)-\tilde{g}_B(p)\tilde{g}_{k_1}(-p)\right)\left(\pi_0\phi_{k_1}-\phi_0\pi_{k_1}\right)\right.\nonumber\\
&&+\left.\pink{2}\tilde{g}_{k_1}(p)\tilde{g}_{k_2}(-p)\left(\phi_{k_1}\pi_{k_2}-\pi_{k_1}\phi_{k_2}\right)
\right]\nonumber\\
&=&\pin{p}p\left[\pink{1}\tilde{g}_B(-p)\tilde{g}_{k_1}(p)\left(\frac{i}{\sqrt{2\omega_{k_1}}}\pi_0(\bd{1}+\bm{1})+\sqrt\frac{\omega_{k_1}}{2}\phi_0(\bd{1}-\bm{1})\right)\right.\nonumber\\
&&+\frac{1}{4}\pink{2}\tilde{g}_{k_1}(p)\tilde{g}_{k_2}(-p)\left(\frac{\omega_{k_1}-\omega_{k_2}}{\sqrt{\omega_{k_1}\omega_{k_2}}}\right)\left(\bd{1}\bd{2}-\bm{1}\bm{2}\right)
\nonumber\\
&&+\left.\frac{1}{4}\pink{2}\tilde{g}_{k_1}(p)\tilde{g}_{k_2}(-p)\left(\frac{\omega_{k_1}+\omega_{k_2}}{\sqrt{\omega_{k_1}\omega_{k_2}}}\right)\left(\bd{1}\bm{2}-\bd{2}\bm{2}\right)
\right]\nonumber.
\eea
As $|0\rangle_0$ is annihilated by $\pi_0$ and $b$ we conclude
\bea
P|0\rangle_0&=&\pink{1}\pin{p}p\tilde{g}_B(-p)\tilde{g}_{k_1}(p)\omega_{k_1}\phi_0\frac{\bd{1}}{\sqrt{2\omega_{k_1}}}|0\rangle_0\\&&+\frac{1}{2}\pink{2}\pin{p}p\tilde{g}_{k_1}(p)\tilde{g}_{k_2}(-p)\left(\omega_{k_1}-\omega_{k_2}\right)\frac{\bd{1}\bd{2}}{\sqrt{4\omega_{k_1}\omega_{k_2}}}|0\rangle_0.
\nonumber
\eea

\subsection{Momentum-Invariant States}

Next, to solve (\ref{princ}) for $|0\rangle_1$, we must first understand how to represent the states in the Hilbert space.  As our operators  $\pi_0$ and $\phi_0$ generate a canonical algebra, they act faithfully on the set of wavefunctions which are functions of $\phi_0$.  The other operators $\bd{i}$ and $b_{k_i}$ generate the $i$th copy of a Heisenberg algebra for a quantum harmonic oscillator.  The corresponding states are products of $\bd{i}$ on $|0\rangle_0$.  As our algebra of operators is the direct sum of the canonical algebra and the oscillator algebras, the states are a tensor product of these representations.  In other words, a general state can be written
\beq
|\psi\rangle=\sum_{m,n=0}^\infty |\psi\rangle_{(n)}^{(m)}\hsp
|\psi\rangle_{(n)}^{(m)}=\pink{n}\psi^{(m)}_{k_1\cdots k_n}(\phi_0)\frac{\bd{1}\cdots\bd{n}}{\sqrt{2^n\omega_{k_1}\cdots\omega_{k_n}}}|0\rangle_0
\eeq
where each $\psi_{k_1\cdots k_n}^{(m)}(\phi_0)$ is a degree $m$ complex polynomial in $\phi_0$.

Noting that $\pi_0$ acts on these wave functions as
\beq
\pi_0\psi^{(m)}_{k_1\cdots k_n}(\phi_0)=\left(-i\frac{\partial}{\partial\phi_0}\psi^{(m)}_{k_1\cdots k_n}(\phi_0)\right)
\eeq
and so 
\beq
\pi_0|\psi\rangle^{(m)}_{(n)}=-i\pink{n}\psi^{(m)\prime}_{k_1\cdots k_n}(\phi_0)\frac{\bd{1}\cdots\bd{n}}{\sqrt{2^n\omega_{k_1}\cdots\omega_{k_n}}}|0\rangle_0
\eeq
we see that the inverse of $\pi_0$ is well-defined up to a $\phi_0$-independent constant of integration $|\psi\rangle^{(0)}$.  Any solution of (\ref{princ}) can therefore be written\footnote{We reserve subscripts in parentheses for counting the number of $b^\dag$, while subscripts of states with no parentheses refer to the semiclassical expansion.}
\bea
|0\rangle_1&=&|0\rangle_1^{(0)}+|0\rangle_1^{(1)}+|0\rangle_1^{(2)}\label{pinv1}\\
|0\rangle_1^{(1)}&=&+\frac{i}{2}\sqrt\frac{\lambda}{8m}\pink{2}\pin{p}p\tilde{g}_{k_1}(p)\tilde{g}_{k_2}(-p)\left(\omega_{k_1}-\omega_{k_2}\right)\phi_0\frac{\bd{1}\bd{2}}{\sqrt{4\omega_{k_1}\omega_{k_2}}}|0\rangle_0
\nonumber\\
|0\rangle_1^{(2)}&=&\frac{i}{2}\sqrt\frac{\lambda}{8m}\pink{1}\pin{p}p\tilde{g}_B(-p)\tilde{g}_{k_1}(p)\omega_{k_1}\phi_0^2\frac{\bd{1}}{\sqrt{2\omega_{k_1}}}|0\rangle_0.\nonumber
\eea
It will be convenient later to remove the inverse Fourier transforms, and so we apply the identities
\beq
\pin{p}p\tilde{g}_B(-p)\tilde{g}_{k_1}(p)=i\int dx g_B(x) g^\prime_k(x)\hsp 
\pin{p}p\tilde{g}_{k_1}(p)\tilde{g}_{k_2}(-p)=i\int dx g^\prime_{k_1}(x) g_{k_2}(x)
\eeq
to obtain
\bea
|0\rangle_1^{(1)}&=&\frac{1}{2}\sqrt\frac{\lambda}{8m}\pink{2}\int dx g^\prime_{k_1}(x) g_{k_2}(x)\left(\omega_{k_2}-\omega_{k_1}\right)\phi_0\frac{\bd{1}\bd{2}}{\sqrt{4\omega_{k_1}\omega_{k_2}}}|0\rangle_0
\nonumber\\
|0\rangle_1^{(2)}&=&-\frac{1}{2}\sqrt\frac{\lambda}{8m}\pink{1}\int dx g_B(x) g^\prime_k(x)\omega_{k_1}\phi^2_0\frac{\bd{1}}{\sqrt{2\omega_{k_1}}}|0\rangle_0.\label{pinv2}
\eea
This is as far as we can get using translation-invariance of the ground state alone.  To determine the $\phi_0$-independent piece, $|0\rangle_1^{(0)}$, we need the Hamiltonian.  That will be the goal of the next section.

\section{The Two Loop Solution} \label{solsez}

To solve the Schrodinger equation (\ref{seqs}) we must apply $H_2$ in (\ref{hfin}) to $|0\rangle_1$, given in Eqs.~(\ref{pinv1}) and (\ref{pinv2}).  

The first term in $H_2|0\rangle_1$ is
\beq
\alpha=\frac{\pi_0^2}{2} |0\rangle_1^{(2)}=\frac{1}{2}\sqrt\frac{\lambda}{8m}\pink{1}\int dx g_B(x) g^\prime_k(x)\omega_{k_1}\frac{\bd{1}}{\sqrt{2\omega_{k_1}}}|0\rangle_0.
\eeq
We will use the equation of motion (\ref{cleq}) together with (\ref{prop1}) and (\ref{prop2}) to make the following manipulations
\bea
\int dx g^\prime_k(x) g_B(x)\omega_k^2&=&-\int dx  \omega_k^2 g_k(x) g^\prime_B(x)\\
&=&-\int dx  \left(\vpp[f(x)] g_k(x)-g_k^{\prime\prime}(x)\right) g^\prime_B(x)\nonumber\\
&=&-\int dx  \left(\vpp[f(x)] g_k(x)  g^\prime_B(x)+g_k^{\prime}(x)g^{\prime\prime}_B(x)\right) \nonumber\\
&=&-\int dx \vpp[f(x)]\left( g_k(x)  g^\prime_B(x)+g^\prime_k(x)  g_B(x)\right) \nonumber\\
&=&-\int dx \vpp[f(x)]\partial_x \left( g_k(x)  g_B(x)\right)\nonumber\\
&=&\int dx \vppp[f(x)]f^\prime(x) g_k(x)  g_B(x) =\sqrt\frac{8m}{\lambda} \vppp_{BBK} \nonumber
\eea
and so we find
\beq
\alpha=\frac{1}{2}\pink{1}\vppp_{BBk_1}\frac{1}{\omega_{k_1}}\frac{\bd{1}}{\sqrt{2\omega_{k_1}}}|0\rangle_0 \label{alft}
\eeq
where we have used the shorthand introduced in Eq.~(\ref{short}).

Similarly the next term is
\bea
\beta&=&\pink{1}\omega_{k_1}\bd{1}b_{k_1} |0\rangle_1^{(2)}=-\frac{1}{2}\sqrt\frac{\lambda}{8m}\pink{1}\int dx g_B(x) g^\prime_k(x)\omega_{k_1}^2 \phi_0^2 \frac{\bd{1}}{\sqrt{2\omega_{k_1}}}|0\rangle_0.\nonumber\\
&=&-\frac{1}{2}\pink{1}\vppp_{BBk_1}\phi_0^2\frac{\bd{1}}{\sqrt{2\omega_{k_1}}}|0\rangle_0=-\pink{1}B_{k_1}\phi_0^2\frac{\bd{1}}{\sqrt{2\omega_{k_1}}}|0\rangle_0.
\eea
We recognize this is as minus the $B_k$ term in $H_3|0\rangle_0$ as written in (\ref{h3form}).  

We have evaluated $(H_2-Q_1) |0\rangle_1^{(2)}$.  Let us now evaluate  $(H_2-Q_1) |0\rangle_1^{(1)}$.  The first term vanishes trivially
\beq
\frac{\pi_0^2}{2} |0\rangle_1^{(1)}=0
\eeq
because $\pi_0|0\rangle_0=0$. The other can be simplified using the identity
\bea
\int dx g^\prime_{k_1}(x) g_{k_2}(x)(\omega_{k_2}^2-\omega_{k_1}^2)&=&\int dx \left( \omega_{k_1}^2 g_{k_1}(x) g^\prime_{k_2}(x)+g^\prime_{k_1}(x) \omega_{k_2}^2g_{k_2}(x)\right)\\
&=&\int dx  \left[\vpp[f(x)]\partial_x\left(g_{k_1}(x) g_{k_2}(x)\right)-\partial_x\left(g^\prime_{k_1}(x) g^\prime_{k_2}(x)\right)\right]\nonumber\\
&=&-\int dx  \vppp[f(x)]f^\prime(x) g_{k_1}(x)  g_{k_2}(x)\nonumber\\
&=&-\sqrt\frac{8m}{\lambda}\vppp_{Bk_1k_2}. \nonumber
\eea
We then find
\bea
\gamma&=&\pink{1}\omega_{k_1}\bd{1}b_{k_1} |0\rangle_1^{(1)}\\
&=&\frac{1}{2}\sqrt\frac{\lambda}{8m}\pink{2}\int dx g^\prime_{k_1}(x) g_{k_2}(x)\left(\omega^2_{k_2}-\omega^2_{k_1}\right)\phi^0\frac{\bd{1}\bd{2}}{\sqrt{4\omega_{k_1}\omega_{k_2}}}|0\rangle_0\nonumber\\
&=&-\frac{1}{2}\pink{2}\vppp_{Bk_1k_2}\phi^0\frac{\bd{1}\bd{2}}{\sqrt{4\omega_{k_1}\omega_{k_2}}}|0\rangle_0=-\pink{2}C_{k_1k_2}\phi_0\frac{\bd{1}\bd{2}}{\sqrt{4\omega_{k_1}\omega_{k_2}}}|0\rangle_0\nonumber
\eea
which again exactly cancels the corresponding term in (\ref{h3form}).

Assembling our results, we have found
\bea
0&=&\left(H_2-Q_1\right) |0\rangle_1+H_3|0\rangle_0\\
&=&\pink{1}\omega_{k_1}\bd{1}b_{k_1} |0\rangle^{(0)}_1+\alpha\nonumber\\
&&+\left(\pink{3}E_{k_1k_2k_3}\frac{\bd{1}\bd{2}\bd{3}}{\sqrt{8\omega_{k_1}\omega_{k_2}\omega_{k_3}}}+\pink{1}F_{k_1}\frac{\bd{1}}{\sqrt{2\omega_{k_1}}}\right)|0\rangle_0\nonumber\\
&=&\pink{1}\omega_{k_1}\bd{1}b_{k_1} |0\rangle^{(0)}_1+\frac{1}{2}\pink{1}\vppp_{BBk_1}\frac{1}{\omega_{k_1}}\frac{\bd{1}}{\sqrt{2\omega_{k_1}}}|0\rangle_0\nonumber\\
&&+\left(\frac{1}{6}\pink{3}\vppp_{k_1k_2k_3}\frac{\bd{1}\bd{2}\bd{3}}{\sqrt{8\omega_{k_1}\omega_{k_2}\omega_{k_3}}}+\frac{1}{2}\pink{1}\int dx \vppp[f(x)]g_{k_1}(x)I(x)\frac{\bd{1}}{\sqrt{2\omega_{k_1}}}\right)|0\rangle_0\nonumber
\eea
where we have used the fact that $\pi_0|0\rangle_1^{(0)}=0$ as $|0\rangle_1^{(0)}$ is independent of $\phi_0$.  This cancellation is critical because, with the $\pi_0^2$ term removed, $H_2$ is invertible and so we can now find $|0\rangle_1$.  To invert $\int \omega b^\dag b$ one need only divide by the sum of the frequencies $\omega$ of each creation operator in the Fock state, yielding
\bea
|0\rangle_1^{(0)}&=&
-\frac{1}{2}\pink{1}\int dx \vppp[f(x)]\frac{g_{k_1}(x)}{\omega_{k_1}}\left(I(x) +\frac{g_B^2(x)}{\omega_{k_1}}\right)\frac{\bd{1}}{\sqrt{2\omega_{k_1}}}|0\rangle_0\nonumber\\
&&-\frac{1}{6}\pink{3}\frac{\vppp_{k_1k_2k_3}}{\omega_{k_1}+\omega_{k_2}+\omega_{k_3}}\frac{\bd{1}\bd{2}\bd{3}}{\sqrt{8\omega_{k_1}\omega_{k_2}\omega_{k_3}}}|0\rangle_0.
\eea
Adding this term to $|0\rangle_1^{(1)}$ and  $|0\rangle_1^{(2)}$ in (\ref{pinv2}) one obtains $|0\rangle_1$, the subleading term in the state $\co|\Omega\rangle$.  This is our main result.  

The most surprising feature is the $g_B^2/\omega$ which is added to the loop factor $I(x)$.  This is the $\alpha$ term from (\ref{alft}).  It is not apparent in expressions for $H_3|0\rangle_0$, but instead is necessary to ensure translation invariance of the soliton ground state $|K\rangle$.  It would be interesting to understand if this correction arises in a diagrammatic approach to the calculation of the ground state.


\section{Remarks}

In general, we do not have a definition of a quantum soliton.  It is a state in the Hilbert space.  We have a definition at zero coupling, where it is a coherent state $\df|\Omega\rangle$ and $f$ is the classical soliton solution.  In the supersymmetric case, if the soliton is BPS, we can follow the soliton to strong coupling by demanding that it remain BPS throughout the deformation.  At weak coupling, we can define a soliton as a Hamiltonian eigenstate given by a semiclassical expansion which starts with the zero coupling state.  That has been the approach in this paper.  The leading quantum correction, corresponding to a squeezed eigenstate of the Poschl-Teller theory, was found in Ref.~\cite{mestato} and the subleading correction $|0\rangle_1$ was found here.  

We believe that the basic strategy employed here, first demanding translation invariance and then solving the Schrodinger equation at the same order, will work to any order in the semiclassical expansion.  But how do we go beyond the semiclassical expansion?   We know in this theory \cite{colemansg} that at strong coupling the soliton becomes the fundamental fermion in the massive Thirring model.  It would be nice to be able to follow it explicitly.  For this, perhaps the low orders in perturbation theory give some hint.  Another possibility would be to consider a supersymmetric version where the soliton is BPS, so that it is described by a first order equation which may be easier to follow.  For this second route, we need to include fermions in our approach.  In this case normal ordering will no longer render the theory finite, and so we need to generalize our formalism to a more general regularization and renormalization prescription.  For example, a Hamiltonian quantization of this system regularized via convolution with a smooth function was introduced in Ref.~\cite{stuart}.  Recently exact supersymmetric coherent states have been constructed in Refs.~\cite{firrotta1,firrotta2}.

Of more immediate concern is the two-loop correction to the Sine-Gordon soliton energy \cite{dhn2loops,luther,vega}.  One expects $\phi_0^2|0\rangle_0$ and $\phi_0^4|0\rangle_0$ terms in both $H_3|0\rangle_1$ and also $H_4|0\rangle_0$.  How is the energy to be extracted from these terms?  In ordinary perturbation theory, one could take the inner product with respect to $|0\rangle_0$ to obtain the energy, but here the $\phi_0$ direction is not normalizable.  Presumably translation invariance will again save us somehow.  In fact, there may be a contribution at the same order from $H_2|0\rangle_2$.  Indeed, invariance under $P\p$ at second order may well lead to a $|0\rangle_2^{(2)}$ and $|0\rangle_2^{(4)}$ term in $|0\rangle_2$.  Perhaps then $H_2|0\rangle_2$ will cancel the unwanted terms from $H_3|0\rangle_1$ and also $H_4|0\rangle_0$?   If there is no such cancellation, one may attack this problem starting with the compactified case \cite{mussardo} where all states are normalizable and so the inner product above is well defined, leading to a direct calculation of the two loop energy.

\section* {Acknowledgement}

\noindent
I thank Hengyuan Guo for a careful reading of this manuscript.  JE is supported by the CAS Key Research Program of Frontier Sciences grant QYZDY-SSW-SLH006 and the NSFC MianShang grants 11875296 and 11675223.   JE also thanks the Recruitment Program of High-end Foreign Experts for support.

\end{document}